\documentclass{llncs}
\usepackage{framed}
\usepackage{url}
\usepackage[ruled,vlined,linesnumbered]{algorithm2e}
\usepackage{comment}
\usepackage{tikz}
\usepackage[utf8]{inputenc}
\usepackage{array}
\usepackage[T1]{fontenc}
\usepackage{todonotes}
\usepackage{float}
\usepackage{amsmath}
\usepackage{multirow}

\usetikzlibrary{arrows,shapes}
\newcommand{\gb}{\textsc{Graph Burning}}
\newcommand{\bg}{\textsc{Burn-Graph}}
\newcommand{\NPH}{\ensuremath{\sf{NP}}-hard\xspace}

\DeclareMathOperator*{\argmax}{argmax}
\DeclareMathOperator*{\argmin}{argmin}

\urldef{\mails}\path|19mcpc06@uohyd.ac.in,askcs@uohyd.ac.in,sdbcs@uohyd.ernet.in|

\newcommand\myprob[3]{%
\begin{framed}
   \par\noindent #1:\\
   {\bfseries Input}: #2\\
   {\bfseries Question}: #3\par
\end{framed}
}

\SetKwComment{Comment}{$\triangleright$\ }{}

\title{Faster Heuristics for Graph Burning}

\author{Rahul Kumar Gautam \and Anjeneya Swami Kare \and S. Durga Bhavani}
\institute{School of Computer and Information Sciences, \\University of Hyderabad,\\ Hyderabad, India\\ \mails}

\begin{document}
\tikzstyle{vertex}=[circle,fill=black!25,minimum size=12pt,inner sep=0pt]
\tikzstyle{selected vertex} = [vertex, fill=red!60]
\tikzstyle{edge} = [draw,thick,-]
\tikzstyle{weight} = [font=\small]
\tikzstyle{selected edge} = [draw,line width=3pt,-,red!50]
\tikzstyle{ignored edge} = [draw,line width=3pt,-,black!20]
\maketitle

 \newcommand{\algoone}[1][]{
\begin{algorithm}[!ht]
\label{algo:greedy}
\DontPrintSemicolon
\SetKwInOut{Input}{Input}\SetKwInOut{Output}{Output}
\Input   {$G=(V,E) $ and a positive integer b}
\Output  {Returns true if algorithm finds a burning sequence of length at most b or false otherwise}
\SetKwFunction{FMain}{BURN-GRAPH}
\SetKwProg{Fn}{Function}{}
\Fn{\FMain{$G$,$b$}} {
\Begin{
   	$G'(V',E') \longleftarrow G(V,E)$\;
	$BS[0 .. b-1]$\Comment*{An array to represent burning sequence}
    	\For{$j \leftarrow 0\; to \; b-1 $}{
		$bestNode \longleftarrow getBestCentralNode(G', b-j-1)$\;
		$BS[j] \longleftarrow bestNode$\;
		$S \longleftarrow N_{G'}^{b-j-1}[bestNode]$\;
		$V' \longleftarrow V' \backslash S$\;
		$G' \longleftarrow G'[V'] $\;
		\If{$ V' = \emptyset $}{
           \For{$k\leftarrow 0\; to\; j$}{
		       print($BS[k]$)
		   }
		   \KwRet true\;		
		}		
	}
    \KwRet false\;
  }
}
\caption{Graph Burning algorithm}
\end{algorithm}
}

\newcommand{\algotwo}[1][]{
  \begin{algorithm}[!ht]
  \label{algo:dijks}
\DontPrintSemicolon
\SetKwInOut{Input}{Input}\SetKwInOut{Output}{Output}
\Input   {$G=(V,E) $}
\Output  {Returns backbone path}
\SetKwFunction{FMain}{Backebone\_path}
\SetKwProg{Fn}{Function}{}
\Fn{\FMain{$G$}} {
\Begin{
		 \ForEach{$v \in  G$}{
		      $ v.\pi \longleftarrow -1 $\;
		      $ v.colour \longleftarrow white $
		 }
		 $ s \longleftarrow min\_central\_node(G) $\;
		 $ s.d \longleftarrow 0 $\;
		 $ s.colour \longleftarrow grey $\;
		 $ s.cs \longleftarrow s.cent $\;
		 $ Q \longleftarrow \phi$\;
		 $ ENQUEUE(Q, s) $\;
		\While{$Q\ne \phi$}{
		   $ u \longleftarrow DEQUEUE(Q) $\;
		
		   \ForEach{$ v \in  G.Adj[u]$}{
		     \If{$v.colour = white$}{
		      $ ENQUEUE(Q, v) $\;
		         $v.colour \longleftarrow grey $ \;
		         $v.d \longleftarrow u.d + 1 $ \;
		         $v.cs \longleftarrow u.cs + v.cent $\;
		         $v.\pi \longleftarrow u$\;
		     }
		     \ElseIf{$v.colour = grey$ and $ v.d > u.d$ and  $v.cs < u.cs + v.cent $}{
		     $v.cs \longleftarrow u.cs + v.cent $\;
		       $v.\pi \longleftarrow u$\;
		     }
		
		   }
		   $u.color \leftarrow black$
		}
	
		\tcc{ Find a node that has heaviest centrality sum with longest distance as a node $nd$.}
		$max\leftarrow -1 $ \;
		$nd\leftarrow nil $\;
		$sm\leftarrow 0 $\;
		  \ForEach{$v \in  G$}{
		     \If{$max <= v.d $}{
		          $max\leftarrow v.d $\;
		          \If{$sm < v.cs $}{
		             $sm\leftarrow v.cs $\;
		             $nd\leftarrow v $\;
		          }
		     }
		  }
		 \tcc{ Return the path as an array $bbpath$  is a dynamic size array.}
		
             	\While{$nd \neq -1 $}{
	            $ append(bbpath,nd) $\;
	            $ nd \leftarrow nd.\pi $\;
	    }
	    \KwRet bbpath \;	
	}
  }
\caption{Backbone path}
\end{algorithm}
}

\newcommand{\algothree}[1][]{
  \begin{algorithm}[!ht]
  \label{algo:bbg}
\DontPrintSemicolon
\SetKwInOut{Input}{Input}\SetKwInOut{Output}{Output}
\Input   {$G=(V,E)$ and a non-negative integer $r$}
\Output  {Returns Best central node}
\SetKwFunction{FMain}{getBestCentralNode}
\SetKwProg{Fn}{Function}{}
\Fn{\FMain{$G, r$}} {
\Begin{
		 $C[0 .. q-1] \longleftarrow getComponents(G)$\;
		 $bbPath[0 .. n-1]$\Comment*{An array to represent vertices in backbone path}
		 $bbLength \leftarrow 0$ \;
         $bbCenrality \leftarrow 0$ \;
		 \ForEach{$comp \in  C$}{
		        $path[0 .. l-1] \leftarrow getBackbonePath(comp)$ \;
		        $pathCentrality \leftarrow \sum_{v\in path}(centrality(v))$ \;
		        \If{$bbLength = l$ and $pathCentrality > bbCenrality $}{
		
		                $bbCenrality \leftarrow pathCentrality$\;
		                $bbpath \leftarrow path$\;
		
		        }
		        \ElseIf{$bbLength < l$}{
                        $bbLength \leftarrow l$  \;
		               $bbCenrality \leftarrow pathCentrality$ \;
		                $bbPath \leftarrow path$\;
		        }
		
		 }
		 $ max \leftarrow 0 $\;
		 $ bestNode \leftarrow null $\;
		\ForEach{$ vertex\;  v \in bbpath  $ in decreasing order of centrality}{
		   $S \leftarrow N_{G}^{r}[bestNode]$\;
		   \If{$|S|>max $ }{
		   $max \leftarrow |S|$\;
		   $bestNode \leftarrow v$\;
		   }
		}
		\KwRet bestNode \;		
	}
  }
\caption{Backbone Based Greedy Heuristic}
\end{algorithm}
}
\newcommand{\algofour}[1][]{
    \begin{algorithm}
    \label{algo:reccomb}
\DontPrintSemicolon
\SetKwInOut{Input}{Input}\SetKwInOut{Output}{Output}
\Input   {$G=(V,E)$, a positive integer $b$ and a dictionary $D$}
\Output  {Returns the length of minimal burning sequence of size at most b or -1 if algorithm fails to find burning sequence of length at most b. Note that the algorithm prints the burning sequence as well.}

\SetKwFunction{FMain}{EstimateBurningNumber}
\SetKwProg{Fn}{Function}{}
\Fn{\FMain{$G$,$b$,$D$}} {
\Begin{
  $bn \longleftarrow -1$\;
  \For{$i \leftarrow b\; down\; to \;1 $}{
    	$G'(V',E') \longleftarrow G(V,E)$\;
	$BS[0 .. b-1]$\Comment*{An array to represent burning sequence}
    	\For{$j \leftarrow 0\; to \; i-1 $}{
		$C[0 .. q-1] \longleftarrow getComponents(G')$\;
		$bestComp \longleftarrow C[0]$\;
		$max \longleftarrow -1 $\;
		$bg \longleftarrow -1 $\;	
		\If{$q > 1$}
		{
			\For{$k\leftarrow 0 \; to \; q-1$}{
				\If{$C[k]\in D$}
				{
					$bg \longleftarrow D[C[k]]$\Comment*{$D$ is a dictionary with component as the key and burning number as the value}
				}
				\Else{
					$bg \longleftarrow {\text{EstimateBurningNumber}(C[k],i,D)}$\;
					$D[C[k]] \longleftarrow bg$\;
				}
				\If{$max < bg$}
				{
					$max \longleftarrow bg$\;
					$bestComp \longleftarrow C[k]$\;
				}
			}
		}
		
		$bbPath \longleftarrow getBackbonePath($bestComp$)$\;
		$ max \longleftarrow 0 $\;
		$ bestNode \longleftarrow null $\;
		\For{$each\; vertex\;  v \in bbpath  $}{
		   $S \longleftarrow N_{bestComp}^{i-j-1}[v]$\;
		   \If{$|S|>max $ }{
		   $max \longleftarrow |S|$\;
		   $bestNode \longleftarrow v$\;
		}
		}
		$BS[j] \longleftarrow bestNode$\;
		$S \longleftarrow N_{bestComp}^{i-j-1}[bestNode]$\;
		$V' \longleftarrow V' \backslash S$\;
		$G' \longleftarrow G'[V'] $\;
		\If{$ V' = \phi $}{
		   $bn \longleftarrow i$\;
		   break\;
		}		
	}
	\If{$V' \neq \phi$}{
		\KwRet bn\;
	}
	
    }
    \KwRet bn\;
  }
  }
\caption{Component Based Recursive Heuristic}
\end{algorithm}
}

\newcommand{\algofive}[1][]{
    \begin{algorithm}[!ht]
    \label{algo:cornheur}
\DontPrintSemicolon
\SetKwInOut{Input}{Input}\SetKwInOut{Output}{Output}
\Input   {$G=(V,E)$ and a non-negative integer $r$}
\Output  {Returns Best central node}
\SetKwFunction{FMain}{getBestCentralNode}
\SetKwProg{Fn}{Function}{}
\Fn{\FMain{$G$,$r$}} {
\Begin{
    $v \longleftarrow \argmax_{v\in V}\{centrality[v]\}$\;
  $S \longleftarrow N_{G}^{r}[v]$\;
  $C[0,1, \cdots q-1] \longleftarrow getComponents(G')$\;
  \If {$q = 0$}
  {
  	 \KwRet $v$\;
  }
  $matrix[][]$\Comment*{A two dimentional array (multi-list) to store vertices in the paths}

  \For{$j \leftarrow 0\; to \; q-1 $}{
  	  $src \longleftarrow \argmin_{u\in C[j]}\{centrality[u]\}$\;
	  $path \longleftarrow getShortestPath(G, v, src)$\;
	  $matrix[j] \longleftarrow path$
  }
  $capacity \longleftarrow |S| $\;
  $bestNode \longleftarrow v $\;
  \For{$ each \; column \;  c \in matrix  $}{
     	$topNodes \longleftarrow getToprNodesByDegree(c) $\Comment*{This function return a list of top $r$ nodes in decreasing order of degree}
	\For{$w \in topNodes$}
	{
		 $S \longleftarrow N_{G}^{r}[w]$\;
		 \If{$capacity \leq |S|$}
		 {
		 	 $capacity \longleftarrow|S|$\;
			 $bestNode \longleftarrow w$\;
		 }
	}
  }

   \KwRet $bestNode$\;
  }
  }
\caption{Improved Cutting Corners Heuristic}
\end{algorithm}
} 
 \newcommand{\picone}[1][]{
\begin{tikzpicture}[scale=1.1]
      \foreach \pos/\name in {{(-1.5,-1)/1}, {(-2,-1)/2}, {(-2,-0.5)/3},
                          {(-2,0)/4}, {(-1.5,0)/5}, {(-1,0)/6},{(-0.5,0)/7}, {(0,0)/8}, {(0.5,0)/9}, {(1,0)/10}, {(1.5,0)/11}, {(2,0)/12}, {(2.5,0)/13}, {(3,0)/14}, {(3.5,0)/15}, {(4,0)/16}, {(3,0.5)/17}, {(3.5,0.5)/18}, {(4,0.5)/19}, {(4.5,0.5)/20}, {(2.5,0.5)/21},{(3.5,-0.5)/22}, {(4,-0.5)/23}, {(4.5,-0.5)/24}, {(2.5,-0.5)/25}, {(3,-0.5)/26}, {(1,0.5)/27}, {(1,1)/28}, {(0.5,1)/29}, {(1,-0.5)/30}, {(1,-1)/31}, {(1.5,-1)/32}, {(0.5,0.5)/33}, {(0,0.5)/34}, {(0,1)/35}, {(0.5,-0.5)/36}, {(0.5,-1)/37}, {(0,-1)/38}, {(1.5,-0.5)/39}, {(2.5,-1)/40}, {(2,-1)/41}, {(1.5,0.5)/42}, {(1.5,1)/43}, {(2,1)/44}, {(0,-0.5)/45}, {(-0.5,-0.5)/46}, {(-1,-0.5)/47}}
       \node[vertex] (\name) at \pos {$\name$};

       \foreach \start /\end in {1/2,2/3,3/4,4/5,5/6,6/7,7/8,8/9,9/10,10/11,11/12,12/13,13/14,14/15,15/17,15/16,15/18,15/19,15/20,15/21,15/22,15/23,15/24,15/25,15/26,10/45,10/36,10/30,10/39,10/33,10/27,10/42,39/40,40/41,30/31,31/32,36/37,37/38,45/46,46/47,33/34,34/35,27/28,28/29,42/43,43/44,35/29,32/41,42/28}
         \path[-,draw,thick] (\start) edge node[name]{} (\end);
\end{tikzpicture}
}

\newcommand{\pictwo}[1][]{
\begin{tikzpicture}[scale=1.1]
      \foreach \pos/\name in {{(-1.5,-1)/1}, {(-2,-1)/2}, {(-2,-0.5)/3},
                          {(-2,0)/4}, {(-1.5,0)/5}, {(-1,0)/6},{(-0.5,0)/7}, {(0,0)/8}, {(0.5,0)/9}, {(1,0)/10}, {(1.5,0)/11}, {(2,0)/12}, {(2.5,0)/13}, {(3,0)/14}, {(3.5,0)/15}, {(4,0)/16}, {(3,0.5)/17}, {(3.5,0.5)/18}, {(4,0.5)/19}, {(4.5,0.5)/20}, {(2.5,0.5)/21},{(3.5,-0.5)/22}, {(4,-0.5)/23}, {(4.5,-0.5)/24}, {(2.5,-0.5)/25}, {(3,-0.5)/26}, {(1,0.5)/27}, {(1,1)/28}, {(0.5,1)/29}, {(1,-0.5)/30}, {(1,-1)/31}, {(1.5,-1)/32}, {(0.5,0.5)/33}, {(0,0.5)/34}, {(0,1)/35}, {(0.5,-0.5)/36}, {(0.5,-1)/37}, {(0,-1)/38}, {(1.5,-0.5)/39}, {(2.5,-1)/40}, {(2,-1)/41}, {(1.5,0.5)/42}, {(1.5,1)/43}, {(2,1)/44}, {(0,-0.5)/45}, {(-0.5,-0.5)/46}, {(-1,-0.5)/47}}
       \node[vertex] (\name) at \pos {$\name$};

       \foreach \pos/\name in{ {(1,0.5)/27},{(1,1)/28}, {(0.5,1)/29}, {(1,-0.5)/30}, {(1,-1)/31}, {(1.5,-1)/32}, {(0.5,0.5)/33}, {(0,0.5)/34}, {(0,1)/35}, {(0.5,-0.5)/36}, {(0.5,-1)/37}, {(0,-1)/38}, {(1.5,-0.5)/39}, {(2.5,-1)/40}, {(2,-1)/41}, {(1.5,0.5)/42}, {(1.5,1)/43}, {(2,1)/44}, {(0,-0.5)/45}, {(-0.5,-0.5)/46}, {(-1,-0.5)/47},{(-0.5,0)/7},{(0,0)/8}, {(0.5,0)/9}, {(1,0)/10}, {(1.5,0)/11}, {(2,0)/12}, {(2.5,0)/13}}
       \node[selected vertex] (\name) at \pos {$\name$};

       \foreach \start /\end in {1/2,2/3,3/4,4/5,5/6,6/7,7/8,8/9,9/10,10/11,11/12,12/13,13/14,14/15,15/17,15/16,15/18,15/19,15/20,15/21,15/22,15/23,15/24,15/25,15/26,10/45,10/36,10/30,10/39,10/33,10/27,10/42,39/40,40/41,30/31,31/32,36/37,37/38,45/46,46/47,33/34,34/35,27/28,28/29,42/43,43/44,35/29,32/41,42/28}
         \path[-,draw,thick] (\start) edge node[name]{} (\end);
\end{tikzpicture}
}
\newcommand{\picthree}[1][]{
    \begin{tikzpicture}[scale=1.1]
      \foreach \pos/\name in {{(-1.5,-1)/1}, {(-2,-1)/2}, {(-2,-0.5)/3},
                          {(-2,0)/4}, {(-1.5,0)/5}, {(-1,0)/6},{(-0.5,0)/7}, {(0,0)/8}, {(0.5,0)/9}, {(1,0)/10}, {(1.5,0)/11}, {(2,0)/12}, {(2.5,0)/13}, {(3,0)/14}, {(3.5,0)/15}, {(4,0)/16}, {(3,0.5)/17}, {(3.5,0.5)/18}, {(4,0.5)/19}, {(4.5,0.5)/20}, {(2.5,0.5)/21},{(3.5,-0.5)/22}, {(4,-0.5)/23}, {(4.5,-0.5)/24}, {(2.5,-0.5)/25}, {(3,-0.5)/26}, {(1,0.5)/27}, {(1,1)/28}, {(0.5,1)/29}, {(1,-0.5)/30}, {(1,-1)/31}, {(1.5,-1)/32}, {(0.5,0.5)/33}, {(0,0.5)/34}, {(0,1)/35}, {(0.5,-0.5)/36}, {(0.5,-1)/37}, {(0,-1)/38}, {(1.5,-0.5)/39}, {(2.5,-1)/40}, {(2,-1)/41}, {(1.5,0.5)/42}, {(1.5,1)/43}, {(2,1)/44}, {(0,-0.5)/45}, {(-0.5,-0.5)/46}, {(-1,-0.5)/47}}
       \node[vertex] (\name) at \pos {$\name$};

       \foreach \pos/\name in{ {(1,0.5)/27},{(1,1)/28}, {(0.5,1)/29}, {(1,-0.5)/30}, {(1,-1)/31}, {(1.5,-1)/32}, {(0.5,0.5)/33}, {(0,0.5)/34}, {(0,1)/35}, {(0.5,-0.5)/36}, {(0.5,-1)/37}, {(0,-1)/38}, {(1.5,-0.5)/39}, {(2.5,-1)/40}, {(2,-1)/41}, {(1.5,0.5)/42}, {(1.5,1)/43}, {(2,1)/44}, {(0,-0.5)/45}, {(-0.5,-0.5)/46}, {(-1,-0.5)/47},{(-1.5,-1)/1}, {(-2,-1)/2}, {(-2,-0.5)/3},{(-2,0)/4}, {(-1.5,0)/5},{(-0.5,0)/7},{(0,0)/8}, {(0.5,0)/9}, {(1,0)/10}, {(1.5,0)/11}, {(2,0)/12}, {(2.5,0)/13}}
       \node[selected vertex] (\name) at \pos {$\name$};

       \foreach \start /\end in {1/2,2/3,3/4,4/5,5/6,6/7,7/8,8/9,9/10,10/11,11/12,12/13,13/14,14/15,15/17,15/16,15/18,15/19,15/20,15/21,15/22,15/23,15/24,15/25,15/26,10/45,10/36,10/30,10/39,10/33,10/27,10/42,39/40,40/41,30/31,31/32,36/37,37/38,45/46,46/47,33/34,34/35,27/28,28/29,42/43,43/44,35/29,32/41,42/28}
         \path[-,draw,thick] (\start) edge node[name]{} (\end);
\end{tikzpicture}
}
\newcommand{\picfour}[1][]{
\begin{tikzpicture}[scale=1.1]
      \foreach \pos/\name in {{(-1.5,-1)/1}, {(-2,-1)/2}, {(-2,-0.5)/3},
                          {(-2,0)/4}, {(-1.5,0)/5}, {(-1,0)/6},{(-0.5,0)/7}, {(0,0)/8}, {(0.5,0)/9}, {(1,0)/10}, {(1.5,0)/11}, {(2,0)/12}, {(2.5,0)/13}, {(3,0)/14}, {(3.5,0)/15}, {(4,0)/16}, {(3,0.5)/17}, {(3.5,0.5)/18}, {(4,0.5)/19}, {(4.5,0.5)/20}, {(2.5,0.5)/21},{(3.5,-0.5)/22}, {(4,-0.5)/23}, {(4.5,-0.5)/24}, {(2.5,-0.5)/25}, {(3,-0.5)/26}, {(1,0.5)/27}, {(1,1)/28}, {(0.5,1)/29}, {(1,-0.5)/30}, {(1,-1)/31}, {(1.5,-1)/32}, {(0.5,0.5)/33}, {(0,0.5)/34}, {(0,1)/35}, {(0.5,-0.5)/36}, {(0.5,-1)/37}, {(0,-1)/38}, {(1.5,-0.5)/39}, {(2.5,-1)/40}, {(2,-1)/41}, {(1.5,0.5)/42}, {(1.5,1)/43}, {(2,1)/44}, {(0,-0.5)/45}, {(-0.5,-0.5)/46}, {(-1,-0.5)/47}}     \node[vertex] (\name) at \pos {$\name$};

       \foreach \pos/\name in{ {(1,0.5)/27},{(1,1)/28}, {(0.5,1)/29}, {(1,-0.5)/30}, {(1,-1)/31}, {(1.5,-1)/32}, {(0.5,0.5)/33}, {(0,0.5)/34}, {(0,1)/35}, {(0.5,-0.5)/36}, {(0.5,-1)/37}, {(0,-1)/38}, {(1.5,-0.5)/39}, {(2.5,-1)/40}, {(2,-1)/41}, {(1.5,0.5)/42}, {(1.5,1)/43}, {(2,1)/44}, {(0,-0.5)/45}, {(-0.5,-0.5)/46}, {(-1,-0.5)/47},{(-1.5,-1)/1}, {(-2,-1)/2}, {(-2,-0.5)/3},{(-2,0)/4}, {(-1.5,0)/5},{(-0.5,0)/7},{(0,0)/8}, {(0.5,0)/9}, {(1,0)/10}, {(1.5,0)/11}, {(2,0)/12}, {(2.5,0)/13},{(3,0)/14}, {(3.5,0)/15}, {(4,0)/16}, {(3,0.5)/17}, {(3.5,0.5)/18}, {(4,0.5)/19}, {(4.5,0.5)/20}, {(2.5,0.5)/21},{(3.5,-0.5)/22}, {(4,-0.5)/23}, {(4.5,-0.5)/24}, {(2.5,-0.5)/25}, {(3,-0.5)/26}}
       \node[selected vertex] (\name) at \pos {$\name$};

       \foreach \start /\end in {1/2,2/3,3/4,4/5,5/6,6/7,7/8,8/9,9/10,10/11,11/12,12/13,13/14,14/15,15/17,15/16,15/18,15/19,15/20,15/21,15/22,15/23,15/24,15/25,15/26,10/45,10/36,10/30,10/39,10/33,10/27,10/42,39/40,40/41,30/31,31/32,36/37,37/38,45/46,46/47,33/34,34/35,27/28,28/29,42/43,43/44,35/29,32/41,42/28}
         \path[-,draw,thick] (\start) edge node[name]{} (\end);
\end{tikzpicture}
}
\newcommand{\picfive}[1][]{
\begin{tikzpicture}[scale=1.1]
      \foreach \pos/\name in {{(-1.5,-1)/1}, {(-2,-1)/2}, {(-2,-0.5)/3},
                          {(-2,0)/4}, {(-1.5,0)/5}, {(-1,0)/6},{(-0.5,0)/7}, {(0,0)/8}, {(0.5,0)/9}, {(1,0)/10}, {(1.5,0)/11}, {(2,0)/12}, {(2.5,0)/13}, {(3,0)/14}, {(3.5,0)/15}, {(4,0)/16}, {(3,0.5)/17}, {(3.5,0.5)/18}, {(4,0.5)/19}, {(4.5,0.5)/20}, {(2.5,0.5)/21},{(3.5,-0.5)/22}, {(4,-0.5)/23}, {(4.5,-0.5)/24}, {(2.5,-0.5)/25}, {(3,-0.5)/26}, {(1,0.5)/27}, {(1,1)/28}, {(0.5,1)/29}, {(1,-0.5)/30}, {(1,-1)/31}, {(1.5,-1)/32}, {(0.5,0.5)/33}, {(0,0.5)/34}, {(0,1)/35}, {(0.5,-0.5)/36}, {(0.5,-1)/37}, {(0,-1)/38}, {(1.5,-0.5)/39}, {(2.5,-1)/40}, {(2,-1)/41}, {(1.5,0.5)/42}, {(1.5,1)/43}, {(2,1)/44}, {(0,-0.5)/45}, {(-0.5,-0.5)/46}, {(-1,-0.5)/47}}
       \node[selected vertex] (\name) at \pos {$\name$};

       \foreach \start /\end in {1/2,2/3,3/4,4/5,5/6,6/7,7/8,8/9,9/10,10/11,11/12,12/13,13/14,14/15,15/17,15/16,15/18,15/19,15/20,15/21,15/22,15/23,15/24,15/25,15/26,10/45,10/36,10/30,10/39,10/33,10/27,10/42,39/40,40/41,30/31,31/32,36/37,37/38,45/46,46/47,33/34,34/35,27/28,28/29,42/43,43/44,35/29,32/41,42/28}
         \path[-,draw,thick] (\start) edge node[name]{} (\end);
\end{tikzpicture}
}

\newcommand{\picsix}[1][]{
\begin{tikzpicture}[scale=1.1]
      \foreach \pos/\name in  {{(-1.5,-1)/1}, {(-2,-1)/2}, {(-2,-0.5)/3},
                          {(-2,0)/4}, {(-1.5,0)/5}, {(-1,0)/6},{(-0.5,0)/7}, {(0,0)/8}, {(0.5,0)/9}, {(1,0)/10}, {(1.5,0)/11}, {(2,0)/12}, {(2.5,0)/13}, {(3,0)/14}, {(3.5,0)/15}, {(4,0)/16}, {(3,0.5)/17}, {(3.5,0.5)/18}, {(4,0.5)/19}, {(4.5,0.5)/20}, {(2.5,0.5)/21},{(3.5,-0.5)/22}, {(4,-0.5)/23}, {(4.5,-0.5)/24}, {(2.5,-0.5)/25}, {(3,-0.5)/26}, {(1,0.5)/27}, {(1,1)/28}, {(0.5,1)/29}, {(1,-0.5)/30}, {(1,-1)/31}, {(1.5,-1)/32}, {(0.5,0.5)/33}, {(0,0.5)/34}, {(0,1)/35}, {(0.5,-0.5)/36}, {(0.5,-1)/37}, {(0,-1)/38}, {(1.5,-0.5)/39}, {(2.5,-1)/40}, {(2,-1)/41}, {(1.5,0.5)/42}, {(1.5,1)/43}, {(2,1)/44}, {(0,-0.5)/45}, {(-0.5,-0.5)/46}, {(-1,-0.5)/47}}
                \node[vertex] (\name) at \pos {$\name$};

       \foreach \pos/\name in{{(1,0.5)/27},{(1,1)/28}, {(0.5,1)/29}, {(1,-0.5)/30}, {(1,-1)/31}, {(1.5,-1)/32}, {(0.5,0.5)/33}, {(0,0.5)/34}, {(0,1)/35}, {(0.5,-0.5)/36}, {(0.5,-1)/37}, {(0,-1)/38}, {(1.5,-0.5)/39}, {(2.5,-1)/40}, {(2,-1)/41}, {(1.5,0.5)/42}, {(1.5,1)/43}, {(2,1)/44}, {(0,-0.5)/45}, {(-0.5,-0.5)/46}, {(-1,-0.5)/47},{(-1,0)/6},{(-0.5,0)/7},{(0,0)/8}, {(0.5,0)/9}, {(1,0)/10}, {(1.5,0)/11}, {(2,0)/12}, {(2.5,0)/13}}
       \node[selected vertex] (\name) at \pos {$\name$};

       \foreach \start /\end in {1/2,2/3,3/4,4/5,5/6,6/7,7/8,8/9,9/10,10/11,11/12,12/13,13/14,14/15,15/17,15/16,15/18,15/19,15/20,15/21,15/22,15/23,15/24,15/25,15/26,10/45,10/36,10/30,10/39,10/33,10/27,10/42,39/40,40/41,30/31,31/32,36/37,37/38,45/46,46/47,33/34,34/35,27/28,28/29,42/43,43/44,35/29,32/41,42/28}
         \path[-,draw,thick] (\start) edge node[name]{} (\end);
\end{tikzpicture}
}
\newcommand{\picseven}[1][]{
\begin{tikzpicture}[scale=1.1]
      \foreach \pos/\name in {{(-1.5,-1)/1}, {(-2,-1)/2}, {(-2,-0.5)/3},
                          {(-2,0)/4}, {(-1.5,0)/5}, {(-1,0)/6},{(-0.5,0)/7}, {(0,0)/8}, {(0.5,0)/9}, {(1,0)/10}, {(1.5,0)/11}, {(2,0)/12}, {(2.5,0)/13}, {(3,0)/14}, {(3.5,0)/15}, {(4,0)/16}, {(3,0.5)/17}, {(3.5,0.5)/18}, {(4,0.5)/19}, {(4.5,0.5)/20}, {(2.5,0.5)/21},{(3.5,-0.5)/22}, {(4,-0.5)/23}, {(4.5,-0.5)/24}, {(2.5,-0.5)/25}, {(3,-0.5)/26}, {(1,0.5)/27}, {(1,1)/28}, {(0.5,1)/29}, {(1,-0.5)/30}, {(1,-1)/31}, {(1.5,-1)/32}, {(0.5,0.5)/33}, {(0,0.5)/34}, {(0,1)/35}, {(0.5,-0.5)/36}, {(0.5,-1)/37}, {(0,-1)/38}, {(1.5,-0.5)/39}, {(2.5,-1)/40}, {(2,-1)/41}, {(1.5,0.5)/42}, {(1.5,1)/43}, {(2,1)/44}, {(0,-0.5)/45}, {(-0.5,-0.5)/46}, {(-1,-0.5)/47}}
       \node[vertex] (\name) at \pos {$\name$};

       \foreach \pos/\name in{{(1,0.5)/27},{(1,1)/28}, {(0.5,1)/29}, {(1,-0.5)/30}, {(1,-1)/31}, {(1.5,-1)/32}, {(0.5,0.5)/33}, {(0,0.5)/34}, {(0,1)/35}, {(0.5,-0.5)/36}, {(0.5,-1)/37}, {(0,-1)/38}, {(1.5,-0.5)/39}, {(2.5,-1)/40}, {(2,-1)/41}, {(1.5,0.5)/42}, {(1.5,1)/43}, {(2,1)/44}, {(0,-0.5)/45}, {(-0.5,-0.5)/46}, {(-1,-0.5)/47},{(-1,0)/6},{(-0.5,0)/7},{(0,0)/8}, {(0.5,0)/9}, {(1,0)/10}, {(1.5,0)/11}, {(2,0)/12}, {(2.5,0)/13},{(2.5,0)/13}, {(3,0)/14}, {(3.5,0)/15}, {(4,0)/16}, {(3,0.5)/17}, {(3.5,0.5)/18}, {(4,0.5)/19}, {(4.5,0.5)/20}, {(2.5,0.5)/21},{(3.5,-0.5)/22}, {(4,-0.5)/23}, {(4.5,-0.5)/24}, {(2.5,-0.5)/25}, {(3,-0.5)/26}}
       \node[selected vertex] (\name) at \pos {$\name$};

       \foreach \start /\end in {1/2,2/3,3/4,4/5,5/6,6/7,7/8,8/9,9/10,10/11,11/12,12/13,13/14,14/15,15/17,15/16,15/18,15/19,15/20,15/21,15/22,15/23,15/24,15/25,15/26,10/45,10/36,10/30,10/39,10/33,10/27,10/42,39/40,40/41,30/31,31/32,36/37,37/38,45/46,46/47,33/34,34/35,27/28,28/29,42/43,43/44,35/29,32/41,42/28}
         \path[-,draw,thick] (\start) edge node[name]{} (\end);
\end{tikzpicture}
}


\newcommand{\graphone}[1][]{
\begin{figure}[hbt]
\centering
\begin{tikzpicture}
      \foreach \pos/\name in {{(-5,5)/1},{(-5,4.5)/2},{(-5,4)/3},{(-5,3.5)/4},{(-5,3)/5},{(-5,2.5)/6},{(-5,2)/7},{(-5,1.5)/8},{(-5,1)/9}, {(-4,4.5)/10},{(-4,4)/11},{(-4,3.5)/12},{(-4,3)/13},{(-4,2.5)/14},{(-4,2)/15},{(-4,1.5)/16},{(-3.5,4.5)/17},{(-3.5,4)/18},{(-3.5,3.5)/19},{(-3.5,3)/20},{(-3.5,2.5)/21},{(-3.5,2)/22},{(-3.5,1.5)/23},{(-3,4.5)/24},{(-3,4)/25},{(-3,3.5)/26},{(-3,3)/27},{(-3,2.5)/28},{(-3,2)/29},{(-3,1.5)/30}}
       \node[vertex] (\name) at \pos {$\name$};

       \foreach \start /\end in {1/2,2/3,3/4,4/5,5/6,6/7,7/8,8/9,10/11,11/12,12/13,13/14,14/15,15/16,17/18,18/19,19/20,20/21,21/22,22/23,24/25,25/26,26/27,27/28,28/29,29/30,20/12,20/26,20/14,20/28,20/13,20/27}
         \path[-,draw,thick] (\start) edge node[name]{} (\end);
\end{tikzpicture}
\caption{An example disconnected graph. The burning number of the graph is $4$.}
  \label{fig:g1}
\end{figure}
}
\newcommand{\graphtwo}[1][]{
\begin{figure}[hbt]
\centering
\begin{tikzpicture}
      \foreach \pos/\name in {{(-3,0)/1}, {(-2.5,0)/2}, {(-2,0)/3},
                          {(-1.5,0)/4}, {(-1,0)/5},{(1,0)/6},{(1,0.5)/7},{(1,-0.5)/8},{(0.5,-0.5)/9},{(0.5,0.5)/10},{(1.5,0.5)/11},{(1.5,-0.5)/12},{(1.5,0)/13},{(0.5,0)/14}}
       \node[vertex] (\name) at \pos {$\name$};
       \foreach \start /\end in {1/2,2/3,3/4,4/5,6/7,6/8,6/9,6/10,6/11,6/12,6/13,6/14,11/13,13/12}
         \path[-,draw,thick] (\start) edge node[name]{} (\end);
\end{tikzpicture}
\caption{An example disconnected graph. The burning number of the graph is $3$.}
  \label{fig:g2}
\end{figure}
}
\newcommand{\graphFirstExample}[1][]{
\begin{figure}[hbt]
\centering
\begin{tikzpicture}
      \foreach \pos/\name in {{(-3,0)/1}, {(-2.5,0)/2}, {(-2,0)/3},
                          {(-1.5,0)/4}, {(-1,0)/5},{(-0.5,0)/6},{(0,0)/7},{(0.5,0)/8},{(-1,-0.5)/10},{(-1,0.5)/12},{(-1.7,0.5)/11},{(-1.7,-0.5)/9}}
       \node[vertex] (\name) at \pos {$\name$};
       \foreach \start /\end in {1/2,2/3,3/4,4/5,6/7,7/8,4/11,4/12,4/9,4/10,3/11,11/12,12/6,3/9,9/10,10/6}
         \path[-,draw,thick] (\start) edge node[name]{} (\end);
\end{tikzpicture}
\caption{An example graph, The vertex sequence $[4, 7, 1]$ is an optimal burning sequence. The burning number of the graph is $3$.}
  \label{fig:bsexample}
\end{figure}
}
\newcommand{\recExample}[1][]{
\begin{figure}[hbt]
\centering
\begin{tikzpicture}
      \foreach \pos/\name in {{(-5,0)/1}, {(-4.5,0)/2}, {(-4,0)/3},
                          {(-3.5,0)/4}, {(-3,0)/5},{(-2.5,0)/6},{(-2,0)/7},{(-1.5,0)/8},{(-1,0)/9},{(-0.5,0)/10},{(0,0)/11},{(0.5,0)/12},{(1,0)/13},{(1.5,0)/14},{(2,0)/15},{(2.5,0)/16},{(3,0)/17},{(3.5,0)/18},{(4,0)/19},{(4.5,0)/20},{(5,0)/21},{(5.5,0)/22},{(6,0)/23},{(6.5,0)/24},{(5.5,0.5)/25},{(6,0.5)/26},{(6.5,0.5)/27},{(1.5,-0.5)/28},{(2,-0.5)/29},{(2.5,-0.5)/30},{(3,-0.5)/31},{(-1,0.5)/32},{(-0.5,0.5)/33},{(0,0.5)/34},{(0.5,0.5)/35},{(-5,0.5)/36}, {(-4.5,0.5)/37},{(-3.5,-0.5)/38},{(-3,-0.5)/39}}
       \node[vertex] (\name) at \pos {$\name$};

     \node[selected vertex] (13) at (1,0) {$13$};
     \node[selected vertex] (21) at (5,0) {$21$};
     \node[selected vertex] (3) at (-4,0) {$3$};
     \node[selected vertex] (7) at (-2,0) {$7$};
     \node[selected vertex] (6) at (-2.5,0) {$6$};
       \foreach \start /\end in {1/2,2/3,3/4,4/5,5/6,6/7,7/8,8/9,9/10,10/11,11/12,12/13,13/14,14/15,15/16,16/17,17/18,18/19,19/20,20/21,21/22,22/23,23/24,25/26,26/27,21/25,28/29,29/30,30/31,32/33,33/34,34/35,36/37,38/39,3/37,3/38,13/35,13/28}
         \path[-,draw,thick] (\start) edge node[name]{} (\end);

\end{tikzpicture}
\caption{An example graph, The vertex sequence $[13, 21, 3,7,6]$ is an optimal burning sequence. The burning number of the graph is $5$.}
  \label{fig:cbrhexample}
\end{figure}
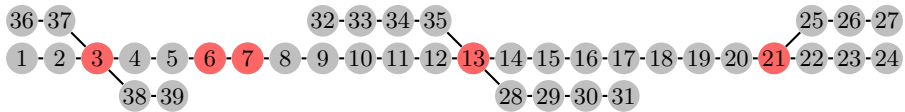
}

\begin{abstract}

Graph burning is a process of information spreading through the network by an agent in discrete steps. The problem is to find an optimal sequence of nodes which have to be given information so that the network is covered in least number of steps. Graph burning problem is  NP-Hard for which two approximation algorithms and a few heuristics have been proposed in the literature. In this work, we propose three heuristics, namely,
Backbone Based Greedy Heuristic (BBGH), Improved Cutting Corners Heuristic (ICCH) and Component Based Recursive Heuristic (CBRH). These are mainly based on Eigenvector centrality measure. BBGH finds a backbone of the network and picks vertex to be burned greedily from the vertices of the backbone. ICCH is a shortest path based heuristic and picks vertex to burn greedily from best central nodes. The burning number problem on disconnected graphs is harder than on the connected graphs. For example, burning number problem is easy on a path where as it is NP-Hard on disjoint paths. In practice, large networks are generally disconnected and moreover even if the input graph is connected, during the burning process the graph among the unburned vertices may be disconnected. For disconnected graphs, ordering of the components is crucial. Our CBRH works well on disconnected graphs as it prioritizes the components.  All the heuristics have been implemented and tested on several bench-mark networks including large networks of size more than $50$K nodes.
The experimentation also includes comparison to the approximation algorithms. The advantages of our algorithms are that they are much simpler to implement and also several orders faster than the heuristics proposed in the literature.
\end{abstract}

\section{Introduction}\label{sec:intro}
 An application of information spread can be seen in public health campaigns. For instance in the case of covid pandemic, the health workers are struggling to get the information reach every house in order to spread awareness and  prevent cornona virus spread.
Take the scenario of one health-worker endeavour. She contacts a family in the village and explains the precautions to be taken up to protect oneself from the contagious infection and persuades that family to help spread  awareness among their acquaintances. We assume that all the members who can be influenced by a family are covered in one step of the awareness campaign. In the burning number context, all those nodes that are covered are called as the 'burning' nodes. Next she has to approach another new family outside this circle of influence to spread the message. The whole process happens in discrete steps. Hence it is important to detect an optimal sequence of families  to be  approached by the health worker that  would ensure coverage of the entire village for propagating the information in minimum time.
In paper~\cite{Simon2019heuristic}, a similar application of passing information during  emergency is discussed. Suppose a piece of  information has to be sent to all the nodes in a network by a  satellite. The satellite informs nodes (hubs) in discrete time steps. All the informed nodes communicate to  their neighbours in parallel. Once a node is informed it is deemed to be in informed state. Process stops when all the nodes are informed.

Graphs are popular representations that model the  social networks of the real  world.  Let $G(V,E)$ be a graph where $V$ is set of nodes depicting people, $E$ denote relationship among the nodes. Initially all nodes are in unburned (uninformed) state.  Given time steps $t_0,t_1,t_2 \cdots t_{b-1}$, at $t_0$ one node is set to fire from outside. It starts burning and spreads the fire to its neighbours in a step wise fashion. 
During the process of burning, it is assumed that either the node is set on fire directly, called as source of fire, or node is burning by catching fire from a neighbour or it is not yet burnt. At $i^{th}$ time step, a new unburned node is set on fire from outside and all those nodes which have caught fire at $t_{i-1}$,  burn their neighbours. The process stops when the entire graph is burning, that is, or all the nodes have received information. Thus the  task is to find the minimum sequence of nodes that have to be chosen as sources of fire, that are directly burned from outside. It is desirable to spread the information through the network, or burn all the nodes of the network  quickly. So the goal is to minimize the number of sources. The minimum number of steps needed to burn the entire graph or the length of the optimal burning sequence is called the \emph{burning number} of the graph. The burning number of the graph $G$ is denoted by $bn(G)$.
Let us consider graph in \figurename~\ref{fig:bsexample}. The vertex sequences $[7, 4, 2, 1]$, $[4, 7, 1]$ and $[3, 6, 8]$ are valid burning sequences, where as the sequence $[7, 4, 1]$ is not a burning sequence. The burning number of the graph is $3$, as the graph does not have burning sequence of length less than $3$.

\graphFirstExample{}

There are a  few problems related to graph burning proposed in the literature. $K$-centre problem~\cite{Garcia2018kcenter}, is one in which the $K$ centres are chosen simultaneously as sources of fire. As a result the nodes burn in parallel and hence very quickly. The K-centres problem is also NP-hard. 
The  Firefighter problem~\cite{Finbow2009firefighting} is a complementary version of the graph burning problem, in which a firefighter protects a node to reduce spread of fire in the graph. At each time step, the firefighter selects a node through which he can protect maximum number of nodes. On the other hand, in graph burning, a node is selected in such a way that it can burn a maximum number of nodes. Therefore the firefighter defends and in graph burning the source of fire burns. Active influence spreading in social networks~\cite{Cordasco2016evangelism,Cordasco2019influence,Kempe2003tss} is the problem of selection of seed sets that can influence as many nodes as possible. 
In this paper, we propose three heuristics for the \gb{} problem. The proposed heuristics are tested on some real world data sets and
their performance is compared to the existing heuristics.

\section{Related Work}
\label{sec:rw}
Bonato et al.~\cite{Bonato2014Contagion} introduced the \gb{} problem and the parameter \emph{burning number}.
They have studied the properties of \gb{} and proposed bounds for burning number. Bessy et al.~\cite{Bessy2017burning}
proved that the decision version of the \gb{} problem is NP-Complete. There has been  lot of attention paid towards studying the \gb{}
problem from theoretical point of view. The complexity and the algorithms for the \gb{} problem for special graph
classes was studied in~\cite{Bonato2016burn,Gupta2020np,Mitche2017probablistics,Kai2017peterson,Bessy2017burning}.
The characterization and bounds for burning number was studied in~\cite{Bessy2018bounds,Bonato2014Contagion,Bonato2016burn,Bonato2019spider,Kamali2020dense,Land2016upper,Lui2019theta,Mitsche2018products}.
Approximation algorithms for the burning number problem was studied in~\cite{Bonato2019approx,Bonato2019spider,Kamali2020dense,Bessy2017burning}.
Parameterized complexity of the \gb{} problem was studied in~\cite{Kare2019parameterized,Kobayashi2020parameters}.

Šimon et al.~\cite{Simon2019heuristic} proposed heuristics for the \gb{} problem. They have studied the
\gb{} problem empirically using both  real world data sets as well as synthetic data sets. They proposed three heuristics based on
Eigenvector centrality, namely Maximum Eigenvector Centrality Heuristic (MECH), Cutting Corners Heuristic (CCH)
and Greedy Algorithm with Forward-Looking Search Strategy Heuristic (GFSSH). The MECH is a greedy heuristic, at each iteration it
selects a (central) node with maximum eigenvector centrality. In CCH, at each iteration, first it finds a set of corner nodes of the graph,
using these corner nodes, a set of central nodes are selected based on eigenvector centrality. Among these central nodes a best central node is
selected using weighted aggregated sum product assessment (WASPAS) algorithm. In GFSSH a set of $20$ central nodes are generated and at each iteration a best central node is selected by combining greedy heuristic and forward looking search.

Šimon et al.~\cite{Simon2019heuristic} also implemented  the $3$-approximation algorithm ($3$-APRX) of Bonato et al.~\cite{Bonato2019approx}
to compare the performance of their heuristics. They have tested the heuristics on some synthetic tree data sets also, however they
did not implement the $2$-approximation algorithm ($2$-APRX) for trees of Bonato et al.~\cite{Bonato2019approx}. As stated in~\cite{Bonato2019approx},
the $2$-approximation algorithm can also be used to compute an upper bound on burning number of any graph. If $G$ is the original graph
and $T$ be any spanning tree of $G$ then $bn(G) \leq bn(T)$. So the $2$-approximation algorithm for trees can also be used to compute
an upper bound on the burning number of the graph. However the computed upper bound need not be a $2$-approximation of the original graph.

Recently, Farokh et al.~\cite{Farokh2020NewHF} proposed six heuristics for the \gb{} problem. Their heuristics are not based on
Eigenvector centrality. They call the vertices in the burning sequence as activators. The first four heuristics are based on different strategies to obtain the first activator and the rest of the activators. First activator is either a central node or it is selected randomly. The rest of the activators are selected such that each vertex has a unique activator. In other words reduce overlapping among the circle around the activators. The other two heuristics are based on diameter, DFS and BFS of the graph.

We propose three heuristics for the \gb{} problem. The heuristics are based on eigenvector centrality. We propose the following heuristics:
Backbone Based Greedy Heuristic (BBGH), Improved Cutting Corners Heuristic (ICCH) and Component Based Recursive Heuristic (CBRH). We have also implemented both $3$-approximation and $2$-approximation algorithms of Bonato et al.~\cite{Bonato2019approx}.
We compare our implemented heuristics with GFSSH, the best performing heuristic of Šimon et al.~\cite{Simon2019heuristic}.
We also compare performance of our algorithms with the results of Farokh et al.~\cite{Farokh2020NewHF}.

Note that both~\cite{Simon2019heuristic} and~\cite{Farokh2020NewHF} tested their heuristics on smaller data sets. We test our heuristics on bigger data sets as well. For example, some of the data sets we used are DIMACS, BOSHLIB, Facebook blue verified pages friends network, DBLP-citation network and huge graphs with size more than $50,000$ nodes like Gemsec-Deezer(HR) (music friendship network in Europe). All three heuristics are performing equally well on all the data sets. Our heuristics are faster than the GFSSH of Šimon et al.~\cite{Simon2019heuristic} and efficient compared to heuristics of Farokh et al.~\cite{Farokh2020NewHF}. Moreover our heuristics are easy to implement.

The rest of the paper is organized as follows: In Section~\ref{sec:heuristics}, we discuss greedy heuristics. The Backbone Based Greedy Heuristic (BBGH) is discussed in Section~\ref{sec:bbpalgo}. The Improved Cutting Corners Heuristic (ICCH) is discussed in Section~\ref{sec:cornerheur}. In Section~\ref{sec:comprecalgo}, we discuss the Component Based Recursive Heuristic (CBRH). In Section~\ref{sec:results}, we discuss our results and some observations. We give conclusions In Section~\ref{sec:concl}.

\section{Proposed Heuristics}
\label{sec:heuristics}
We consider the following decision version of the \gb{}, which asks to check if the graph $G$ can be burned in at most $b$ time steps. If we have an algorithm  for \bg{$(G, b)$}, we can use binary search to compute minimum $b$ for which the \bg{$(G, b)$} returns true. As the \gb{} is \NPH{}, we can not expect to have a exact polynomial time algorithm for \bg{$(G, b)$}. In this paper, we propose heuristics for \bg{$(G, b)$}. The heuristics for \bg{$(G, b)$}, if it returns $true$, it means that the algorithm is successful in finding a burning sequence of length at most $b$. If it returns $false$, it means that the algorithm failed to find a burning sequence of length at most $b$. Note that, in the later case, the graph can still be burned in at most $b$ steps.

\myprob{\bg{$(G, b)$}}{A graph $G = (V,E)$ and a positive integer $b$.}{Does the graph have a burning sequence of length at most $b$?}

We propose three heuristics for \bg{$(G, b)$}. First two are greedy in nature and the third one is a component based recursive algorithm. First we discuss the greedy algorithms and then the recursive algorithm.

\par{}
At the outset, the process underlying the greedy algorithms is as shown in Algorithm~\ref{algo:greedy}. At each iteration, the function $getBestCentralNode()$ extracts an unburned vertex to burn next. The two heuristics differ in the procedure $getBestCentralNode()$.

\algoone{}

\subsection{Backbone Based Greedy Heuristic (BBGH)}
\label{sec:bbpalgo}
We call a longest path starting at a node with minimum centrality value and containing nodes with high centrality values as the backbone path. A node on the backbone path can potentially burn more vertices. With this intuition we propose the Backbone Based Greedy Heuristic (BBGH). We extract the backbone path and for each vertex in the backbone path, we see how many vertices the vertex can burn, if we burn the vertex in the current time step. The vertex which leads to maximum number of burned vertices is chosen as the best central node. Here the crucial step is: how to extract the backbone path?.

\algothree{}

A backbone path is a longest path starting at  a node with minimum centrality and containing nodes with high centrality values. To compute backbone path we use BFS traversal. We compute BFS tree rooted at a node with minimum centrality value. In this rooted tree we look at the nodes at highest depth, there can be more than one such node. For all these nodes we consider shortest path from the root to the node and compute average centrality  of all the nodes in the path. A path with maximum average centrality is considered as the backbone path. The procedure $getBackbonePath()$  returns a path.

\begin{table}
\caption{Trace of BBGH, the Algorithms~\ref{algo:greedy} and~\ref{algo:bbg}. For each iteration vertices shown in red color are not part of the graph $G$. The estimated burning number of the graph is $4$.}
\label{table:bbg}
\begin{centering}
\begin{tabular}{|m{4cm}|>{\centering\arraybackslash}m{9cm}|}
\hline
 Original Graph $(G)$ & \picone{}\\
 \hline
 \multicolumn{2}{|m{13cm}|}{$1^{st}$ Iteration: Graph is connected and hence it has a single component. Backbone path $bbpath=[16, 15, 14, 13, 12, 11, 10, 9, 8, 7, 6, 5, 4, 3, 2, 1]$ and best central node is $10$ for radius $3$. Here, by radius we mean the number of remaining steps in the burning sequence.)}\\
 \hline
 After $1^{st}$ Iteration & \pictwo{} \\
 \hline
 \multicolumn{2}{|m{13cm}|}{$2^{nd}$ Iteration: Graph has two components and backbone path $bbpath=[6, 5, 4, 3, 2, 1]$ and best central node is $3$ for radius $2$.}\\
 \hline
 After $2^{nd}$ Iteration & \picthree{} \\
 \hline
 \multicolumn{2}{|m{13cm}|}{$3^{rd}$ Iteration: Graph has two components and backbone path $bbpath=[16, 15, 14]$ and best central node is $15$ for radius $1$.}\\
\hline
  After $3^{rd}$ Iteration &  \picfour{} \\
 \hline
 \multicolumn{2}{|m{13cm}|}{$4^{th}$ Iteration: Graph has one component and backbone path $bbpath=[6]$ and best central node is $6$ for radius $0$.}\\
 \hline
 After $4^{th}$ Iteration &  \picfive{} \\
 \hline
 \end{tabular}
  \end{centering}
\end{table}

Let us take backbone path as an array $bbPath$. For each vertex $v \in bbPath$ in decreasing order of centrality values, compute $S = N_{G'}^{r}[v]$, set of all the vertices which are at a distance at most $r$ from $v$. Then whichever node gives maximum $|S|$ value, will become the best central node. If the graph is disconnected, then backbone path of each component is extracted and which ever vertex of these backbone paths gives maximum $|S|$ value, we return that vertex as the best central node. The complete algorithm is shown in Algorithm~\ref{algo:bbg}. Working of the BBGH is shown in Table~\ref{table:bbg} with an example. Note that, for the graph given in Table~\ref{table:bbg},  GFSSH of Šimon et al.~\cite{Simon2019heuristic} gives a burning sequence of size $5$, where as our BBGH burns the graph in $4$ time steps.

\subsection{Improved Cutting Corners Heuristic (ICCH)}

\label{sec:cornerheur}
Šimon et al.~\cite{Simon2019heuristic}, presented heuristic called Cutting Corner Heuristic (CCH). Their algorithm has $O(mn)$ time complexity. We present a similar heuristic which also runs in worst case $O(mn)$  time. However our algorithm is easy to implement and runs faster in practice as we avoid computation of average path length and call to weighted aggregated sum product assessment (WASPAS) method.

Let $r$ be the number of time steps available to burn the graph. We start by computing the centrality values. Let $u$ be the node with maximum
centrality. We remove the $r$ neighborhood of $u$, that is, all the vertices in the set $N_{G}^{r}[u]$ from the graph $G$. If the resulting graph is empty, then we return the vertex $u$ as the best central node. Otherwise, let the resulting graph have $q$ components, say $C[0], C[1], \cdots, C[q-1]$. For each component $C[i]$, $0 \leq i < q$, we take the minimum centrality vertex (say $v_i$) of the component $C[i]$ and compute shortest path from $u$ to $v_i$ and let the shortest path be denoted by $P[i]$. We visualize $P[0], P[1], \cdots, P[q-1]$ as a matrix (multi-list), where each $P[i]$ is treated as a row in the matrix. Now for each column of the matrix, we pick $r$ nodes in decreasing order of degree. If $c$ is the number of columns of the matrix, we will get at most $r*c$ such nodes. For each of these nodes, we compute $S = N_{G}^{r}[.]$ value. Then whichever node gives maximum $|S|$ value, will become the best central node. The process is depicted in the Algorithm~\ref{algo:cornheur}. Working of the ICCH is shown in Table~\ref{table:corner} with an example. From our results, we observe that our ICCH performs better than the CCH of Šimon et al.~\cite{Simon2019heuristic}.

\newpage
\algofive{}

\begin{table}[!ht]
\caption{Trace of ICCH, the Algorithms~\ref{algo:greedy} and~\ref{algo:cornheur}. For each iteration vertices shown in red color are not part of the graph $G$. The estimated burning number of the graph is $5$.}
\label{table:corner}
\begin{centering}
\begin{tabular}{|m{4cm}|>{\centering\arraybackslash}m{9cm}|}
\hline
 Original Graph $(G)$ & \picone{}\\
 \hline
 \multicolumn{2}{|m{13cm}|}{$1^{st}$ Iteration: graph is connected and hence it has a single component. We get node $10$ as the next node to burn for radius $4$.}\\
 \hline
 After $1^{st}$ Iteration & \picsix{} \\
 \hline
 \multicolumn{2}{|m{13cm}|}{$2^{nd}$ Iteration: We get node $15$ as the next node to burn for radius $3$.}\\
 \hline
 After $2^{nd}$ Iteration & \picseven{} \\
 \hline
 \multicolumn{2}{|m{13cm}|}{$3^{rd}$ Iteration: We get node $3$ as the next node to burn for radius $2$.}\\
\hline
  After $3^{rd}$ Iteration &  \picfive{} \\
 \hline
  \end{tabular}
 \end{centering}
\end{table}

\subsection{Component Based Recursive Heuristic (CBRH)}
\label{sec:comprecalgo}

If the graph is disconnected, choosing of a component to burn a vertex can make a difference. Let us consider the graphs in \figurename~\ref{fig:g1} and \figurename~\ref{fig:g2}. These graphs have two components. Let us see the following criterion to select the component to burn a vertex.

 \begin{enumerate}
 \item{\textbf{Component with maximum size: } If we choose component with maximum number of vertices, we need a burning sequence of length $4$ to burn the graph in \figurename~\ref{fig:g1} and we need a burning sequence of length $4$ to burn the graph in \figurename~\ref{fig:g2}. But the actual burning number of the graph in \figurename~\ref{fig:g2} is $3$}.
 \item{\textbf{Component with maximum path length: } If we choose component with maximum path length, we need a burning sequence of length $5$ to burn the graph in \figurename~\ref{fig:g1} and we need a burning sequence of length $3$ to burn the graph in \figurename~\ref{fig:g2}. But the actual burning number of the graph in \figurename~\ref{fig:g1} is $4$.}
 \end {enumerate}
\graphone{}
\graphtwo{}
\recExample{}
For the graph given in \figurename~\ref{fig:g1}, the Backbone Based Greedy Heuristic will choose the component with vertex set $\{1,2,3 \cdots,9\}$ and vertex $5$ is chosen as the best central node. Now we can see that we require $5$ time steps  to burn the graph. But the burning number of the graph is $4$. The optimum burning sequence chooses a vertex $20$ from the second component. Note that even if graph is connected in the beginning, in the subsequent iterations the graph can be disconnected and this situation can arise. Therefore choosing a component is very crucial. Therefore the question is what is the criteria to select the best component of the graph. Ideally we should choose a component that has maximum burning number.

In the Component Based Recursive algorithm, we recursively run the Backbone Based Greedy Heuristic and which ever component leads to maximum burning number, we choose such a component. For each component the algorithm recursively estimates the burning number of the component. During the recursive calls, burning number computed for a component is stored in a dictionary to avoid redundant recursive calls. The component with maximum estimated burning number is selected and from which best central node is selected. The process is described in the Algorithm~\ref{algo:reccomb}. For the example considered in Table~\ref{table:bbg}, the trace of the algorithm is very similar that of Algorithm~\ref{algo:bbg}.

Our BBGH and ICCH fails to compute optimal burning sequence of either graph in \figurename~\ref{fig:g1} or \figurename~\ref{fig:g2}. But our CBRH computes optimal burning sequence for both the graphs. For the connected graph given in \figurename~\ref{fig:cbrhexample}, our CBRH computes the optimal burning number but our other two heuristics, BBGH, ICCH and  GFSSH of Šimon et al.~\cite{Simon2019heuristic} fail to compute optimal burning sequence. This concludes the significance of our ICCH.

\algofour{}

 \section{Results and Discussion}
 \label{sec:results}
 We tested our heuristics on the following data sets:
 \begin{enumerate}
   \item The Network Data Repository~\cite{Rossi2015data}
    \begin{itemize}
     \item Netscience
     \item Polblogs
     \item Reed98
     \item Mahindas
     \item Cite-DBLP
   \end{itemize}
   \item Stanford large network dataset collection (SNAP Datasets)~\cite{2014SnapData}
    \begin{itemize}
     \item Chameleon
     \item TVshow
     \item Ego-Facebook
     \item Squirrel
     \item Politician
     \item Government
     \item Crocodile
     \item Gemsec-Deezer(HR)
   \end{itemize}
 \end{enumerate}
 We also generated $100$ random trees and tested the performance of the algorithms. All the heuristics are implemented in Python programming language. The algorithms have been implemented on a system with processor Intel Core $i5$, processor speed of $2.7$ GHz having dual core and $8$GB RAM.

 We compare performance of our heuristics with those of ~\cite{Simon2019heuristic} and~\cite{Farokh2020NewHF}. The Table~\ref{table:simcomp} shows the estimated values of the burning number for various algorithms. We have compared our results with $3$-approximation and $2$-approximation algorithms of Bonato et al.~\cite{Bonato2019approx}, GFSSH, the best performing heuristic of~\cite{Simon2019heuristic}\footnote{We requested their code and run on our machine.}. We have also listed the number of recursive calls made by our Component Based Recursive Heuristic. It can be observed that, at the outset the algorithm looks like an exponential algorithm, but in practice the number of recursive calls made is very less even for bigger graphs. Note that for all the social networking data sets and other data sets that have been considered in this paper, the diameter (radius) of the graph is very small. The burning number of a graph with radius $r$ is at most $r+1$. As the radius of the graphs is very small, improving the burning by even a small number is tough. From Table~\ref{table:simcomp} we observe that our heuristics are competitive to the best heuristic of Šimon et. al.~\cite{Simon2019heuristic}.  Table~\ref{table:time} shows the running time comparison of our heuristics with that of GFSSH, the best performing heuristic of~\cite{Simon2019heuristic}. Our heuristics are faster than that of ~\cite{Simon2019heuristic}.
 Note that both Šimon et. al.~\cite{Simon2019heuristic} and Farokh et al.~\cite{Farokh2020NewHF} tested their heuristics on smaller data sets. As our heuristics are faster, they take lesser time on even bigger data sets.

 While comparing with heuristics of Farokh et al.~\cite{Farokh2020NewHF}, it can be seen that our heuristics give better results for some of the graphs. The Table~\ref{table:farcomp} shows the comparison of performance of our heuristics with that of~\cite{Farokh2020NewHF}.



\begin{table}[!ht]
\scriptsize
\centering
\caption{Comparison of estimated burning number of approximation algorithms~\cite{Bonato2019approx}, GFSSH of~\cite{Simon2019heuristic} and our heuristics.  The last column shows the number of recursive calls made by CBRH. *-For Netscience and Mahindas data sets,~\cite{Simon2019heuristic} quoted burning number as $6$, however when we run their program on our machine we got $7$ and $5$ respectively.}
\begin{tabular}{| m{2.0cm} | m{2.5cm} | >{\centering\arraybackslash} m{1cm} |>{\centering\arraybackslash}m{0.8cm} |>{\centering\arraybackslash}m{0.8cm} | >{\centering\arraybackslash} m{0.5cm}| >{\color{red}} >{\centering\arraybackslash}  m{0.5cm}  |>{\color{blue}} >{\centering\arraybackslash} m{0.5cm}|>{\centering\arraybackslash}  m{0.5cm}|>{\centering\arraybackslash}m{0.5cm}| >{\centering\arraybackslash}m{0.6cm}|>{\centering\arraybackslash}m{0.6cm}|}
\hline
  \textbf{Network Source } &\textbf{Name  }  &  \textbf{|V|}  &\textbf{|E|}&\rotatebox{90}{\textbf{3-APRX}} &\rotatebox{90}{\textbf{2-APRX}} & \rotatebox{90}{ \textbf{GFSSH~\cite{Simon2019heuristic}}}&
 \rotatebox{90}{{\textbf{BBGH}}} & \rotatebox{90}{\textbf{ICCH}}& \rotatebox{90}{\textbf{CBRH}} &\rotatebox{90}{\textbf{CBRH Calls}}  \\
 \hline
 \multirow{ 5}{*}{\parbox{2.0cm}{Network Data Repository}}&{Netscience  } &379 & 914&12& 10  & 7* & 7 & 7 & 7 & 23 \\
 \cline{2-11}
 &{Polblogs } &{643} & {2K}& 9 &  10 & 6 &  6 &  6 & 6  & 8 \\
 \cline{2-11}
 &{Reed98  } & {962} & {18K }& 6 &8  & 4  & 4 & 4& 4 &46\\
\cline{2-11}
 &{Mahindas} & 1258 & 7513& 9 & 8  & 5* & 5 & 5 & 5& 68 \\
 \cline{2-11}
  &{Cite-DBLP} & 12.6K & 49.7K & 120 & 82  & 41 & 41 & 41 &  41 &146\\
 \hline
 \multirow{ 8}{*}{SNAP Data set }
 &{Chameleon} &  2.2K& 31.4K & 9 &  10  & 6 & 6 & 6 &6 &43 \\
 \cline{2-11}
 &{ TVshow} &3.8K&17.2K& 18&  16 &  10  & 10 & 10 &  10 & 49\\
 \cline{2-11}
  &{Ego-Facebook  } &  {4K} & {88K} & 9 & 6  &4  & 4  & 4 & 4  & 110\\
 \cline{2-11}
 &{Squirrel } &  5K & 198K& 9&  10  &6 &  6 & 6 &  6 & 19\\
 \cline{2-11}
  &{Politician} & 5.9K & 41.7K&  12 & 12  & 7 & 7 &  7 &  7 & 9\\
 \cline{2-11}
 &{Government} &7K& 89.4K& 9 &  10  & 6 &  6 &  6 &  6 & 7\\
 \cline{2-11}
  &{Crocodile  } &  11K &170K & 12&  10& 6 & 6 &  6&  6 & 40\\
 \cline{2-11}
 &{ Gemsec-Deezer(HR)} & 54K & 498K& 12& 12 & 7 &7 &  7 & 7 & 92  \\
 \hline
  \multirow{ 2}{*}{\parbox{2.0cm}{Randomly \text{}generated}} &{Barabasi-Albert} &  {1K} & {3K} &6 & 8  & 4.9 & 4.9 &  4.9 &4.9  & 8\\
 \cline{2-11}
 &{Erdos-Renyi } &  1K & 6K& 6& 8  & 5 &  5 &  5 & 5 & 1 \\
 \hline
\end{tabular}
\label{table:simcomp}
\end{table}

\begin{table}[!ht]
\scriptsize
\centering
\caption{Comparison of running times of our heuristics with GFSSH of~\cite{Simon2019heuristic} }

\begin{tabular}{|m{2.0cm} |m{2.5cm} |>{\centering\arraybackslash} m{1cm} |>{\centering\arraybackslash}m{1cm} |>{\centering\arraybackslash}m{1cm} | >{\centering\arraybackslash}m{1cm}| >{\centering\arraybackslash}m{1cm}  |>{\centering\arraybackslash}m{1cm}|>{\centering\arraybackslash}m{1cm}|}

\hline
 \textbf{Network Source } &\textbf{Name  }  &  \textbf{|V|}  &\textbf{|E|}& \textbf{ GFSSH~\cite{Simon2019heuristic}}&
 \textbf{BBGH} & \textbf{CBRH}& {\textbf{ICCH}} \\
 \hline
 \multirow{5}{*}{\parbox{2.0cm}{Network Data Repository}}&{Netscience  } &379 & 914 &2m& <1s& 1s& \color{red} <1s\\
 \cline{2-8}
   &{Polblogs } &  {643} & {2K} &3s & \color{red} <1s  & 2s & \color{red} 1s \\
 \cline{2-8}
 &{Reed98  } & {962} & {18K } &  5s  & 3s&  5s& \color{red} 3s \\
 \cline{2-8}
  &{Mahindas} & 1258 & 7513 & 6s & \color{red} <1s & 23s & 3s \\
 \cline{2-8}
  &{Cite-DBLP} & 12.6K & 49.7K & 3m 8s & 39s & 2m & \color{red} 22s \\
 \hline
 \multirow{ 8}{*}{SNAP Data set } &{Chameleon} & 2.2K& 31.4K &25s& \color{red} 8s&20s&16s \\
  \cline{2-8}
  &{TVshow} & 3.8K&17.2K& 30s & \color{red} 7s&15s&22s\\
  \cline{2-8}
  &{Ego-Facebook  } & {4K} & {88K} & 1m & 17s & 22s& \color{red} 16s\\
  \cline{2-8}
 &{Squirrel } & 5K & 198K & 3m 5s&40s&1m 40s& \color{red}34s \\
  \cline{2-8}
 &{Politician} &5.9K & 41.7K& 1m& \color{red} 14s& 17s& 32s \\
  \cline{2-8}
   &{Government} & 7K& 89.4K & 1m 13s&\color{red} 20s& 32s&50s\\
  \cline{2-8}
  &{Crocodile  } &  11K &170K&5m&  2m
 36s& 4m&\color{red} 42s \\
  \cline{2-8}
  &{Gemsec-Deezer(HR)} & 54K & 498K & 1h 20m& \color{red} 2m 36s & 47m&7m \\
 \hline
 \multirow{ 2}{*}{\parbox{2.0cm}{Randomly \text{}generated}}&{Barabasi-Albert } & {1K} & {3K} &2m & \color{red} 1s& 2s& \color{red} 1s \\
  \cline{2-8}
 &{Erdos-Renyi } &  1K & 6K & 6m& 1s&3s& \color{red} <1s \\
 \hline

\end{tabular}
\label{table:time}
\end{table}

\begin{table}[!ht]
\scriptsize
\centering
\caption{Comparison of estimated burning number of heuristics of~\cite{Farokh2020NewHF} and our heuristics.}
\begin{tabular}{|m{1.4cm}|>{\centering\arraybackslash}m{0.6cm}|>{\centering\arraybackslash}m{0.8cm}|m{0.5cm}|>{\centering\arraybackslash}m{0.5cm}|>{\centering\arraybackslash}m{0.4cm}|>{\centering\arraybackslash}m{0.4cm}|>{\centering\arraybackslash}m{0.4cm}|>{\centering\arraybackslash}m{0.4cm}|>{\centering\arraybackslash}m{0.4cm}|>{\centering\arraybackslash}m{0.4cm}|>{\centering\arraybackslash}m{0.4cm}|>{\centering\arraybackslash}m{0.4cm}|>{\centering\arraybackslash}m{0.4cm}|>{\centering\arraybackslash}m{0.4cm}|}
\hline
\textbf{Name} & \textbf{|V|} & \textbf{|E|} & \textbf{Max. deg} & \textbf{Avg. deg} & \rotatebox{90}{\textbf{ctr-Half}} & \rotatebox{90}{\textbf{ctr-far}} & \rotatebox{90}{\textbf{Rand-Half} }& \rotatebox{90}{\textbf{Rnd-Far}} & \rotatebox{90}{\textbf{DFS-path}} &\rotatebox{90}{\textbf{D-BFS-Path}} &\rotatebox{90}{\textbf{BBGH}}&\rotatebox{90}{\textbf{CBRH}}&\rotatebox{90}{\textbf{ICCH}}  \\
\hline
c-fat200-1 & 200 &1534 & 17 & 15 & 11 & 8 & 9&\color{blue} 7 & 8 &8 & \color{red} 7 & 7 & 7 \\
\hline
c-fat200-2 & 200 & 3235 & 34 & 32 & 6 & 6 & 6  & 5 & \color{blue} 5 &  6 & \color{red} 5 & 5 &5\\
\hline
c-fat200-5 & 200 & 8473 & 86 & 84 & 4 & 4 & 4 & 4 & 4 & \color{blue}3 & \color{red} 3& 3 & 3\\
\hline
c-fat500-1 & 500 & 4459 & 20 & 17 & 12 & 11 & 12 & \color{blue} 10 & 15 & 17 & \color{red} 9 & 9 & 10\\
\hline
c-fat500-10 & 500 & 46627 & 188 & 186 & 4 & 4 & 4 & \color{blue} 4 & 4 & 4 & \color{red} 3 &3 &3 \\
\hline
c-fat500-2 & 500 & 9139 & 38 & 36 & 9 & 8 & 9 & 8 &  11 & \color{blue} 8 & \color{red} 7 &7 & 7 \\
\hline
c-fat500-5 & 500 & 23191 & 95 & 92 & 6 & 6 & 6 & 6 & \color{blue} 5 & 6 & \color{red} 5 & 5 & 5 \\
\hline
\end{tabular}
\label{table:farcomp}
\end{table}

\section{Conclusion}
\label{sec:concl}

In this paper, we proposed three heuristics for \gb{} problem, namely, Backbone Based Greedy Heuristic (BBGH), Improved Cutting Corners Heuristic (ICCH) and Component Based Recursive Heuristic (CBRH). Firstly, we show a need for each heuristic by constructing the required example graphs. That is, we show a graph for which   BBGH finds a better burning number compared to GFSSH; a graph on which ICCH finds a better burning number than BBGH and finally need for the recursive algorithm of CBRH, where CBRH manages to find a better burning number than the other heuristics. 

We show through extensive experimentation that BBGH turns out to be the fastest among all the heuristics and several orders faster than the heuristic GFSSH of Šimon et al. which is one of the latest heuristics proposed for this problem as shown in Table~\ref{table:time}. To give an example, on the largest network of the benchmark data set with  size $54$K, BBGH gave burning number of $7$ in  $2m$ $36s$ as compared to $1h$ $20m$ taken by GFSSH. ICCH follows as a close second by delivering the result in $7$ minutes. 
Lastly, the recursive heuristic though is slower than our other heuristics   it runs faster than GFSSH on most of the large data sets.  It shows a way to prioritize the ordering of selection of components in order to obtain optimal burning number. Further, we also show the superior results obtained by our proposed heuristics on other data sets experimented by Farokh et al. in Table~\ref{table:farcomp}.

We feel that the techniques used in the paper can be extended to active influence spreading problems like, Target Set Selection, Perfect Seed Set and Perfect Evangelic Set problems.

\bibliographystyle{splncs03}
\bibliography{myrefs}
\end{document}